\documentclass[12pt,journal,communications surveys and tutorials]{IEEEtran}
\usepackage{epstopdf}
\usepackage[pdftex]{graphicx}
\usepackage{subfig}

\ifCLASSOPTIONcompsoc
\else
\fi
\ifCLASSINFOpdf
\else
\fi
\begin{document}
%
\title{A Survey Report on Operating Systems for Tiny Networked Sensors}
%
%
%
%

\author{Alok~Ranjan,
        H.B.~Sahu and~Prasant.~Misra
\IEEEcompsocitemizethanks{ \IEEEcompsocthanksitem A.Ranjan and H.B.Sahu are with the Department of Mining Engineering, Natioanl Institute of Technology, Rourkela,Odisha
 769008, India. 
E-mail: alokranjan007@hotmail.com
\IEEEcompsocthanksitem P. Misra is with Robert Bosch Center for  Cyber Physical Systems, Indian Institute of Science, Bangalore, Karnataka, India.}
\thanks{}}

\IEEEtitleabstractindextext{%
\begin{abstract}
Wireless sensor network (WSN) has attracted researchers worldwide to explore the research opportunities, with application mainly in health monitoring, industry automation, battlefields, home automation and environmental monitoring. A WSN is highly resource constrained in terms of energy, computation and memory. WSNs deployment ranges from the normal working environment up to hostile and hazardous environment such as in volcano monitoring and underground mines. These characteristics of WSNs hold additional set of challenges in front of the operating system designer. The objective of this survey is to highlight the features and weakness of the opearting system available for WSNs, with the focus on the current application demands. The paper also discusses the operating system design issues in terms of architecture, programming model, scheduling and memory management and support for real time applications.
\end{abstract}

}

\maketitle

\IEEEdisplaynontitleabstractindextext

%
\IEEEpeerreviewmaketitle

\section{Introduction}
%
%

%
%
%
%
\IEEEPARstart{W}
ith the distinguished communication features, WSNs are becoming a critical infrastructure support in health monitoring, environmental monitoring, tracking, industrial automation, volcano monitoring, indoor application, and etc. \cite{Akyildiz2002,chehri2010experimental,misra2010safety,1607983,yick2008wireless}.
Recent innovations in the field of wireless sensor network technology are significantly enhancing the capabilities for distributed sensing of the physical world. A wireless sensor network consists of a number of sensor nodes deployed densely that are equipped with own data processing, communication and sensing capabilities. The advantage of having a WSN is low cost, increased coverage and importantly the operating capabilities in a normal environment up to hostile environments like in underground mines, underwater, under soil and also those unattended areas where manual monitoring is tough such as volcano monitoring \cite{misra2010safety,1607983,yick2008wireless,ranjan2014advancements}.
An operating system is there in every node to execute the network protocols for communication with other nodes in the network. The OS is responsible for managing resources on every sensor node and provides a programming interface to the system developer. It's an OS only which makes an application program portable and simple by providing a level of abstraction to the hardware. The OS for sensor node is different than the general purpose computers in terms of the processor architecture, sensor devices and hardware for communication \cite{dutta2012operating}. An operating system mainly manages system resources with the multiplexing concept in two ways, i.e. space (memory) and time. A space multiplexing incorporates various programs accessing parts of the the available resource, whereas the time multiplexing involves programs turn in using the system resources \cite{phani2007operating}. In this paper, we have examined the core features of the operating systems for WSNs in terms of programing model, memory management, architecture and protocols for communication. We also discuss the future research issues related to communication protocols at each layer. This survey paper focuses on those OSs which are widely used in the recent scenario in both academic and industrial research. The remainder of this paper is organized as follows. In section 2, we briefly discuss the sensor node architecture followed by the open research issues of the communication protocols. Different sensor nodes widely used in academic research community and commercial purposes are also discussed in brief with a comparative  study on those motes. Section 3, discuss the popular OSs for WSN. Open research issues with the OS design consideration is discussed in section 4. Section 5 covers the future remarks based on our studies. We then conclude our work in section 6.

\section{Sensor Node Architecture}
Typically, a sensor has two parts: a sensing element which senses the physical environment and a transducer that converts the sensed data to a representative signal. Nodes typically have on board memory, sensors, wireless connectivity, power source and on board processing capabilities \cite{sammarco2007technology}. Major components of a sensor node include controller, memory, sensors/actuators, communication module and power and shown in figure 1. A controller is responsible for processing all the relevant data and is also capable of execution of arbitrary code. Memory is used mainly to store some programs and intermediate data. Sensors and actuators are the elementary part of the node and is the interface to the physical world for sensing physical environment and control physical parameters. Sending and receiving is done through the communication part. Power is most significant and critical part of the node and is very constrained for WSN node. The beauty of the sensor node fitted with an on board is, the node transmits only the required and necessary data instead of sending raw data to a node which is responsible for data fusion. They use their own computation and processing capabilities \cite{Akyildiz2002}. The overall energy consumption of each sensor node consists of three components, viz. the energy consumed in sensing, i.e. sensing energy, the amount of energy consumed in data processing and energy required for communication between nodes \cite{chen2005chain}. Where sensing energy solely related to the sensitivity of the sensors, data processing energy related to the circuit design and the energy consumed in communication is almost half of the energy consumed by each node. In addition to communication process, it has also been found that the energy consumptions in communication process have also three components: radio energy, transmitter device electronic energy and receiver node electronics energy \cite{zhou2008reliable}. For any kind of network, connectivity is one of the fundamental issues for reliable network functionalities.
\begin{figure}[t]
\centering
\includegraphics[width=0.7\linewidth,height=0.3\textheight]{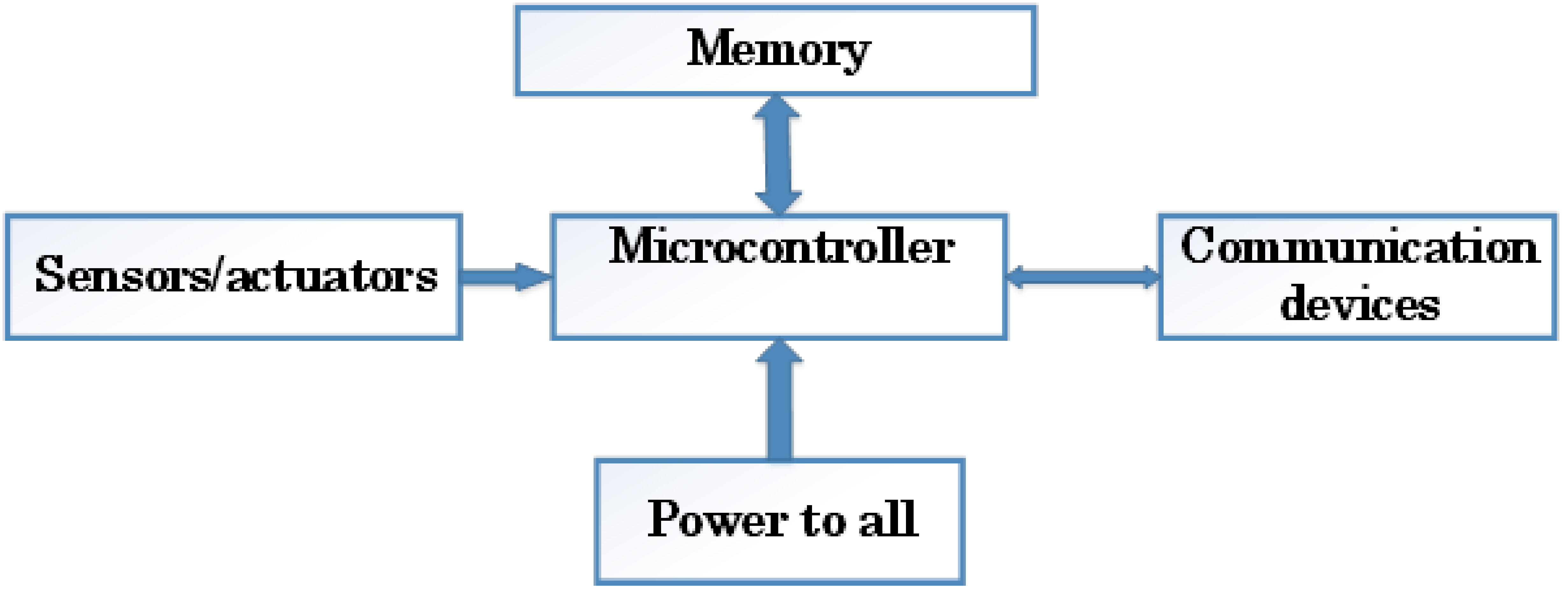}
\caption{Simplified architecture of sensor node}
\label{fig:alok4-eps-converted-to}
\end{figure}

\subsection{Communication Protocols and issues}
A sensor network consists of different layers in a protocol stack. Physical layer deals with transmission media, modulation technique, receiving of bits and connectivity among nodes. Due to the dynamic topology of a sensor node Medium access control (MAC) applied to a sensor network must be aware of power consumption and should be able to minimize the collision of data packets \cite{yick2008wireless}. Routing of data is taken care by the network layer and it further gives services to the transport layer. End to end data flow into a sensor network is with transport layer functionalities. Various power efficient algorithms have been proposed which is smart enough so that after receiving a message the node may turn off its receiver, which prevents the further receipt of duplicate messages. Additionally, a node can go into sleep mode for a specified amount of time and can pass a broadcast message in the network about shorting in power so can’t participate in routing \cite{yick2008wireless,akyildiz2002wireless}. This gives an obvious power saving and this can further utilize in sensing. Although there are a number of literatures available describing the different layer functionalities of WSN. The main objective of this section is to address some open research issues. This section highlights only some critical research issues at every layer so that the reliability and efficiency of communication devices can be addressed in the future research.
\begin{figure}[b]
\centering
\includegraphics[width=0.7\linewidth, height=0.25\textheight]{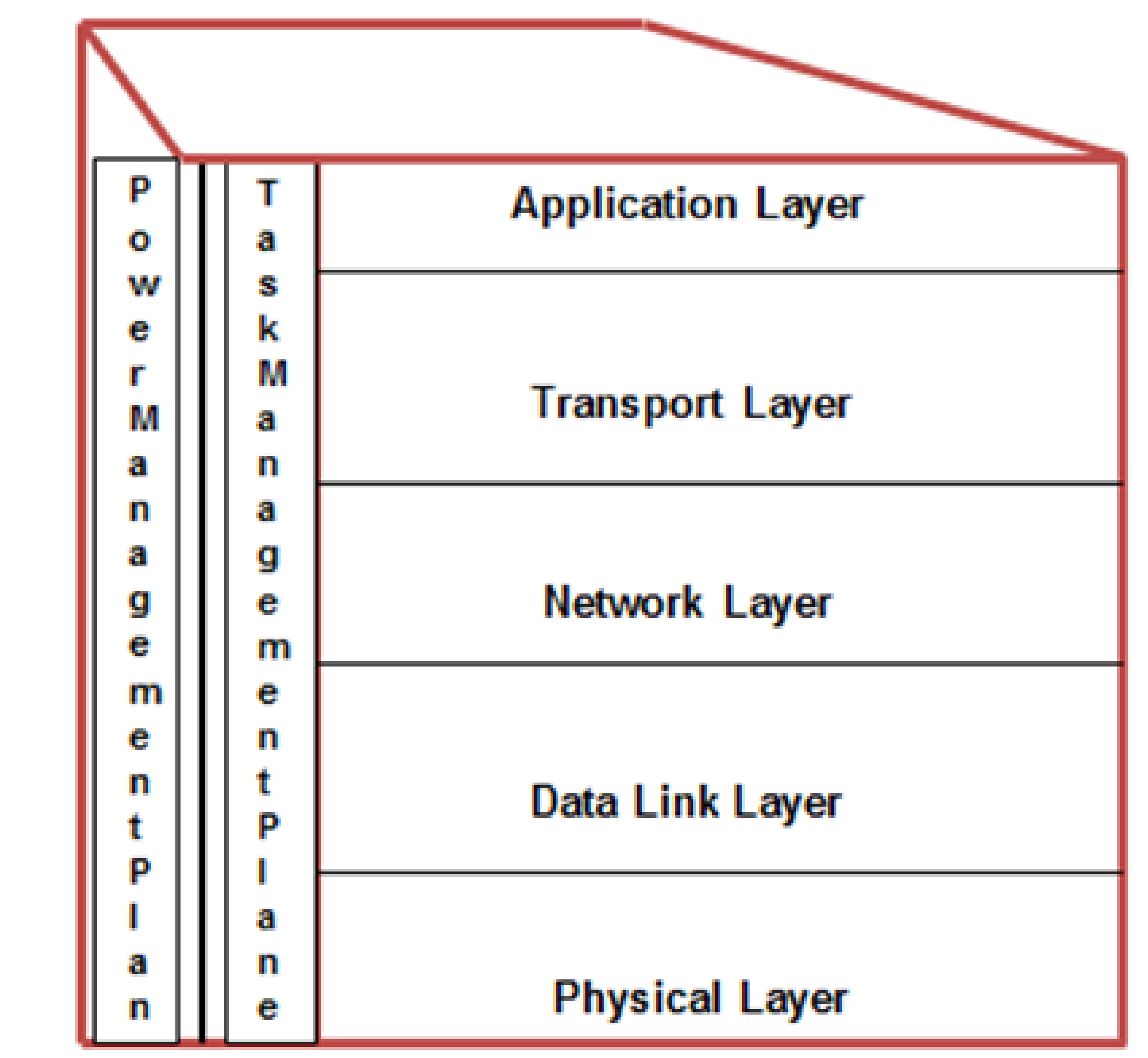}
\caption{WSN Communication Protocol Architecture}
\label{fig:alok5-eps-converted-to}
\end{figure}

\subsubsection{Physical layer}
Physical layer deals with the transmission of data, connectivity, modulation techniques and synchronizations. This faces different challenges for wireless communication to be possible. This layer interacts with the MAC layer of data link layer performing reception and bit synchronization. In \cite{akyildiz2006wireless}, the authors found that the error rate at the physical layer is high when the distance between transmitter and receiver is increased. Energy consumption is one of the critical point for WSN. As the WSN has limited power source, hence to minimize the energy consumption and maximizing the life of a sensor node is vital. Physical layer uses power for bit stream transmission and radio operation. The amount of energy consumes to transmit a data stream can be adversely affected and varies due to the interference, multipath effects, fading and distance between the transmitter and receiver \cite{akyildiz2006wireless}. Whereas the energy consumption is fixed to operate circuitry. The energy consumption of the different radio chip used in WSN nodes is listed in subsection "Sensor Nodes". An efficient modulation technique for channel is another issue which is important for minimizing energy consumption by the network. \cite{shih2001physical,wang2001energy} have carried out comparative analysis between binary modulation and multi-level (M-ary) modulation scheme and noticed that the energy consumed by the M-ary modulation is less and energy efficient than binary modulation. This occurs when the output power of RF is small and the time interval for start-up is less. On further comparison they found that when M is greater than eight, the M- ary phase shift keying and M-ary quadrature amplitude modulation scheme consumes more energy than the M-ary frequency shift keying. In addition to energy consumption, a comparative analysis of Binary and M-ary has been done and it was found that  M-ary is capable to reduce the transmit time by transmitting multiple bits per symbol but it leads to a complex circuitry which in results consumes more power \cite{daniels1996surface}. Further in their studies they suggested that binary modulation is good because it takes less power hence, is energy efficient. Choices for bandwidth also needs research efforts, as in WSN there are three major choices available at physical layer: spread spectrum, narrow band and ultra wideband (UWB). It may be noted that research efforts are there by the researchers on communication possibilities based on spread spectrum and narrow band, however, there are only limited literatures available which have the information on UWB based sensor communications. Therefore, the scope of UWB based sensor communications  required more research efforts and implementations in real world scenarios.

\subsubsection{Data Link Layer}
The Data link layer is responsible for data transfer between the two; not sharing the same link. For better functionalities the MAC layer protocol design should have the following features \cite{akyildiz2002survey,yick2008wireless}: support scalability, maximize bandwidth utilization, data frame synchronization and energy efficient. Transport layer and data link layer both deals with the error detection and correction mechanisms. There are issues for designing MAC protocols which include mainly network topology, energy consumption and throughput of the overall network.  A number of researchers have proposed solutions to address energy consumptions in sensor nodes. The various power saving modes and mechanism for listening network to minimize energy consumption by the nodes have been reported by different researchers which helps to enhance the lifetime of the network \cite{guo2001low,mishra2004adaptive,rajendran2006energy}. Cross layer optimization and overall system performance are one of the future issues which can significantly improve the performance of the WSN. Dynamic change in topology and mobile network has forced researchers to explore the optimal design considerations for future needs of industries. The number of transmission must be minimized for maximizing the life of sensor nodes. In WSN there are various techniques used for recovery of data includes simple packet combining (SPAC)\cite{dubois2005packet}, automatic repeat request (ARQ) \cite{ganti2006datalink}, forward error correction (FEC)\cite{blahut1983theory}, hybrid ARQ (HARQ) \cite{liu1997error} and multi radio diversity (MRD) \cite{miu2005improving}. Among these all FEC, significantly decreases the number of data transmission. This gives the advantage of reduced acknowledgement wait time and also the retransmission of data is avoided. Efficient research on trade-offs between the contention based MAC protocol and time divison multiple access (TDMA) /carrier sense multiple access (CSMA) is required \cite{yick2008wireless}. To target WSN communication applications in industries, Forward Error Check (FEC) technique is required to be researched extensively so that the retransmission of data can be minimized.

\subsubsection{Network Layer}
A network layer protocol design for WSN should address the critical issues of fault tolerance, scalability, and efficiency \cite{Akyildiz2002}. Network protocol is needed for sensor nodes to build a communication network. The network protocol plays an important role to define the sequential instructions for the sensor nodes to take in order to communicate with other nodes and also for data formats.  A sensor network is further challenged by the physical communication environment followed by the parameters like bit error rate, dynamic topology and also the resources available to individual nodes. Routing protocols used in an ad-hoc network is categorized into three: reactive, proactive and geographical. Reactive protocols establish a route discovery whenever required in the network. While the proactive routing protocols are responsible for monitoring paths between the devices in the network. Routes are setup based on the physical location of the devices by the geographical routing protocols \cite{akyildiz2006wireless}.  Research on low duty cycles of the network should be examined. Routing protocols satisfying Quality of Service (QOS) can surely minimize end to end delay and energy efficiency. Research efforts are needed for QOS routing protocols in wireless sensor network. Design of routing protocol should withstand with the continuous changing topology of the network and must support scalability for future needs \cite{yick2008wireless}. It should be such that the network performance may not get  affected.

\subsubsection{Transport Layer}
Reliability and data accuracy at the receiver and sender ends are taken care by the transport layer protocols. The design of transport layer protocol should be such that it supports the concept generic i.e. for all applications \cite{yick2008wireless}. As the different communication devices are developed by different manufacturers, the transport layer protocols should support heterogeneous network. Therefore, the developed protocols should have the vision of universal acceptance and can be applied to any kind of communication devices.  Any loss in data packets directly affects the quality of service and wasted energy. Therefore, the protocol should be smart enough which can minimize this issue and able to detect and recover the lost data. It surely helps for better throughput and low energy consumptions. Reliability of network is directly connected to the throughput. Optimization of sensed data at nodes is another issue to reduce the congestion in the network.  Trade-off between the loss of information and delay is further research issue for WSN. 

\subsubsection{Cross Layer Interaction}
 MAC layer capabilities of cross layer communications can save energy with the routing knowledge of neighbors. The variations in power level for transmission can be utilized further by the routing protocols. It is done with the knowledge that which link is taking the least amount of power to transmit. This effort reduces the energy consumption and hence, increases the lifetime of the network. Cross layer designs for WSN enhance the performance of the network. Changing topology due to the mobility can be useful for other nodes to transmit and update the route information. A cross layer MAC design shares this topology information in the network to support reliable routing and maximizing coverage for communication \cite{akyildiz2006wireless,chehri2012empirical}.
\subsection{SENSOR NODES}
Practically WSN nodes or motes vary in size ranging from a diameter of less than 1 cm disc shaped boards to enclosed systems having dimensions of 5 cm square \cite{johnson2009comparative}. A node in the WSN has mainly two objectives: (a) it is used both for data processing i.e. sensed data from the physical environment and forward it to sink or other nodes; sometimes it also works as a gateway node  and (b) Data logging. There are different sensor nodes available in the market for commercial purpose and academic research works. We list the comparative analysis on the basis of specifications and features of different sensor nodes studied in \cite{johnson2009comparative} in table 1 and figure 3.
\paragraph{TeleosB/ Tmote Sky}
This mote is developed by UC Berkeley and widely used in academic researchers and also in some industry specific applications. It is currently available from CrossBow and Sentilla technology.
\paragraph{Mica2/MicaZ}
This is another sensor node popularly used. This is the product of CrossBow Technology and belongs to the second and third generation wireless sensor networking mote. It is possible to connect additional sensor boards like humidity sensors, temperature sensors and so on to the node. Connections between the sensor and the controller are depending upon the version of the nodes (Mica2, MicaZ, Mica). SPI and 12C bus are frequently used for connections.
\paragraph{EYES}
“Energy Efficient Sensor Networks” were the sponsored project of European Union. Infineon designed the nodes and it is equipped with the MSP430 microcontroller proprietary of Texas Instrument. The node has a USB interface to communicate with PC and can attach with additional sensor boards.
\paragraph{SHIMMER}
This mote is famous for the wearable kind of applications like in smart jackets, health monitoring applications. SHIMMER stands for “Sensing Health with Intelligence, Modularity, Mobility and Experimental Reusability. It is currently available from Real Time Ltd.
\paragraph{SUN SPOT}
SUN SPOT is a WSN mote manufactured by Sun Microsystems and stands for “Small Programmable Object Technology”. For this both software and hardware are open source and easily available.
\paragraph{EZ430-RF2480/2500}
This is from Texas Instruments incorporated with the MSP430 microprocessor and having CC2480/2500 radio transceiver on board. These motes are very economical in use.
\begin{table}%
 \caption{Comparative analysis of different sensor nodes\label{tab:four}}
 {\includegraphics[scale=.06]{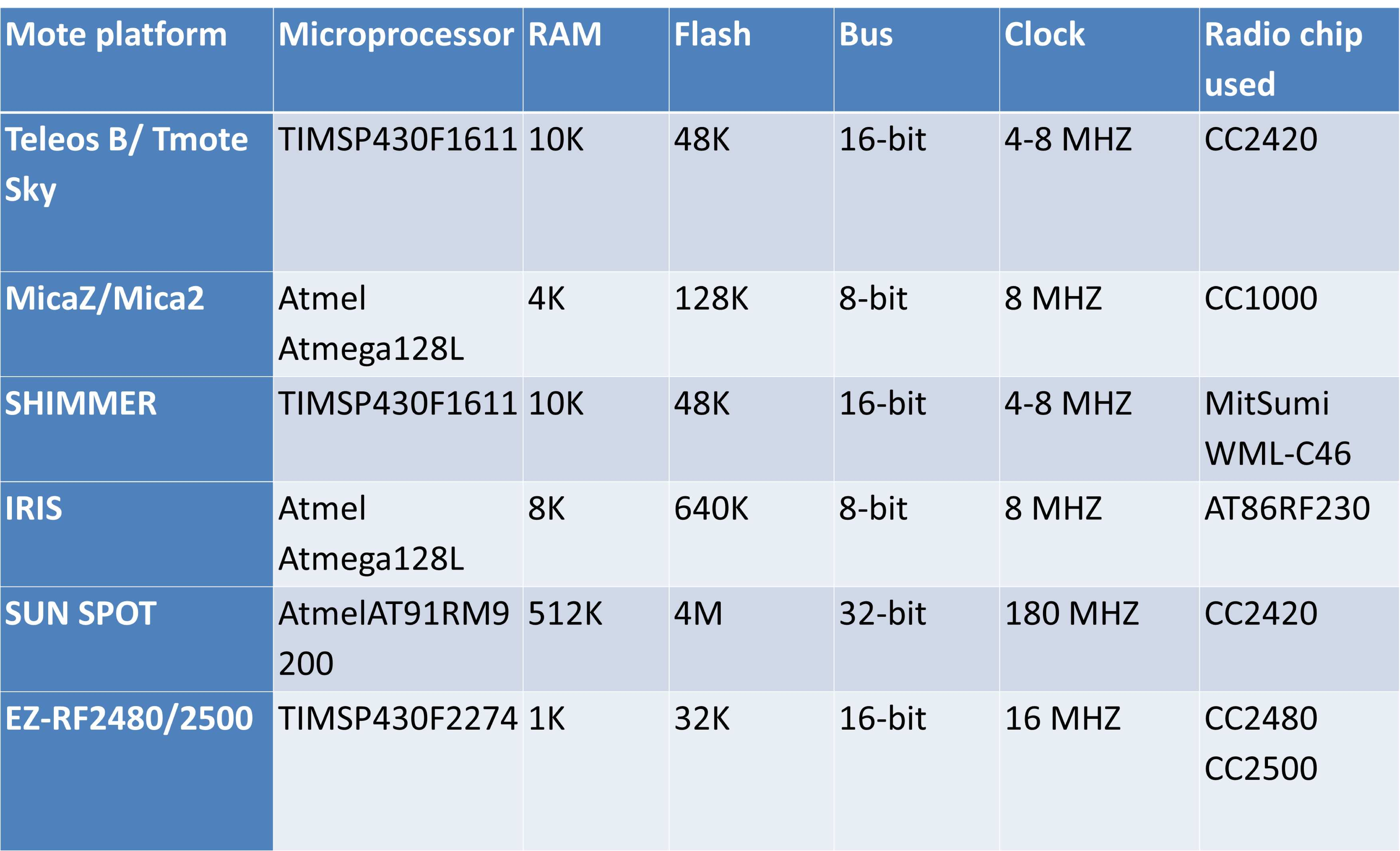}}
 \end{table}
 \begin{figure*}[t]
\centering
\includegraphics[width=0.7\linewidth, height=0.4\textheight]{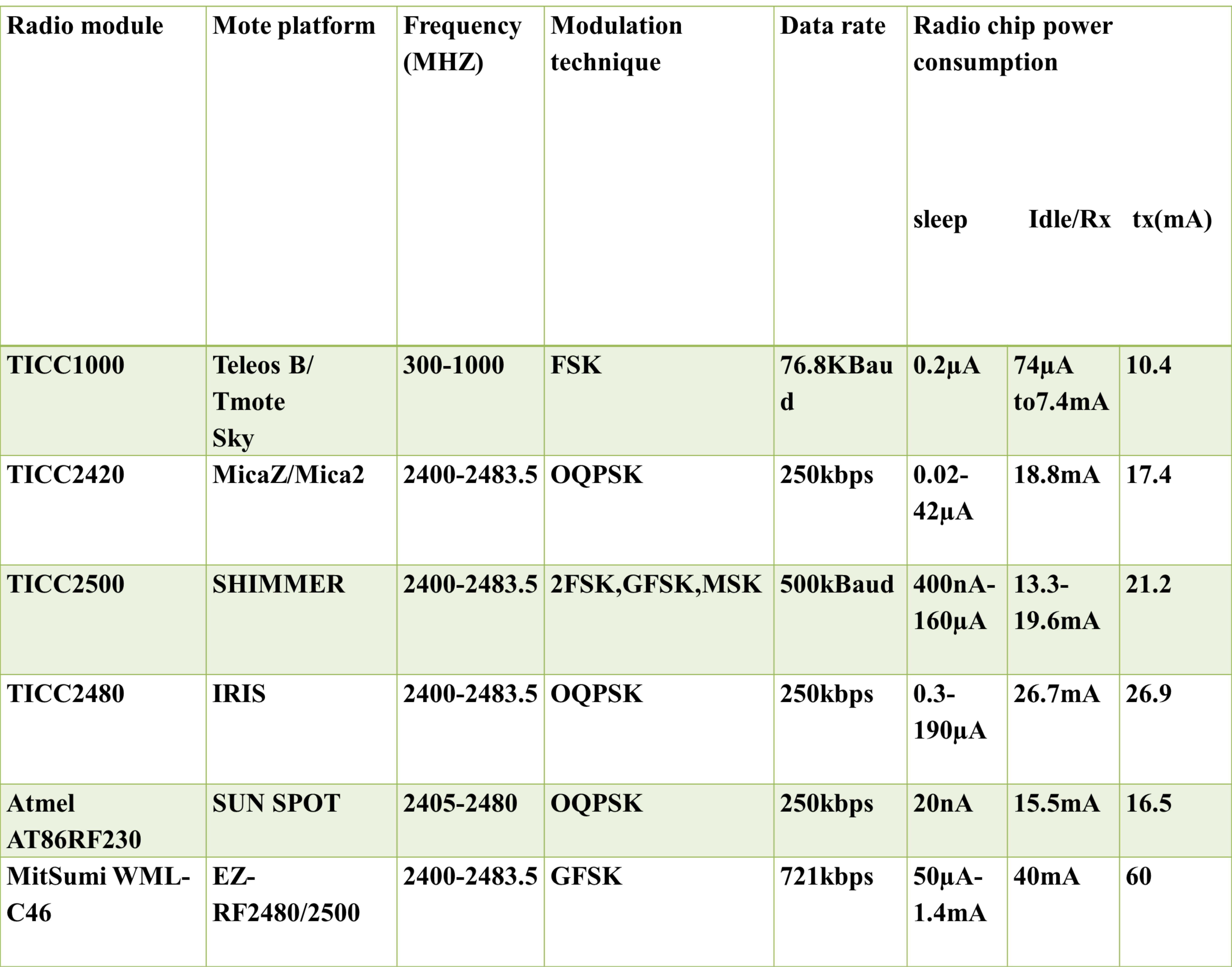}
\caption{Specifications of different radio chips used by sensor motes platform}
\label{fig:table9-eps-converted-to}
\end{figure*}
\section{Operating Systems for WSN}
The nature of wireless sensor networks is highly dynamic because the nodes in the network are seriously affected by either due to different environmental conditions or power consumptions by the network. A WSN is composed of sensor nodes, which is very resource constrained in terms of energy, memory and processing. Optimization of life of a sensor node is a fundamental challenge because in most of the scenario WSN is deployed where it is not possible to attend the nodes and also to replace a sensor node after deployment is complex \cite{farooq2011operating}. Compared to traditional, contemporary processing units a microcontroller used in WSN node operates at very low frequency and having limited processing capabilities. To address these kind of above said issues, the need for a smart operating system is desired \cite{yick2008wireless}. The OS always acts as a resource manager and its main objective is to maximize resource utilizations. An OS is responsible for resource allocation among users and it is further achieved by managing users in a controlled and orderly manner. An OS provides the interface to the physical world by providing an abstraction to the sensor node.  Characteristics of WSNs impose a set of additional challenges to the OS designer. The design of OS should be such that it supports the maximum resource utilization in an optimized way.   The following coming section covers widely used OS for embedded devices and WSNs in academic researches and commercial applications as well. We further summarize a comparative study based on features of different OSs in figure 6, which are popularly known and is based on the studies \cite{farooq2011operating,phani2009operating}.
\subsection{TinyOS}
TinyOS \cite{levis2005tinyos}, it is an open source and flexible OS and perhaps the first operating system designed mainly for resource constrained devices such as WSN. It follows the monolithic architecture design. This OS is based on component based approach and uses NesC , a C dialect programming language. Figure 4, shows a simplified TinyOS architecture. The footprint of TinyOS is 400 bytes therefore it is suitable for low memory requirements. Early release of TinyOS had no support for multithreading, but with the later version 2.1 of TinyOS it supports multithreading.  These threads supported by the OS are called TOS threads. TinyOS has active message communication protocol support. It does not support dynamic memory management; it has only static memory management support. TinyOS manages resource sharing using two mechanisms: Virtualization and Events completion. A virtualization mechanism deals resources as an independent instance and the application which uses resource is independently operated. An event completion mechanism for resource sharing is used when the resources are not managed by virtualization mechanism. The size of the foot print of TinyOS is small as compared to the other well-known OS for embedded systems and WSNs such as Contiki.\nocite{levis2012experiences}
\begin{figure}[t]
\centering
\includegraphics[width=0.7\linewidth, height=0.25\textheight]{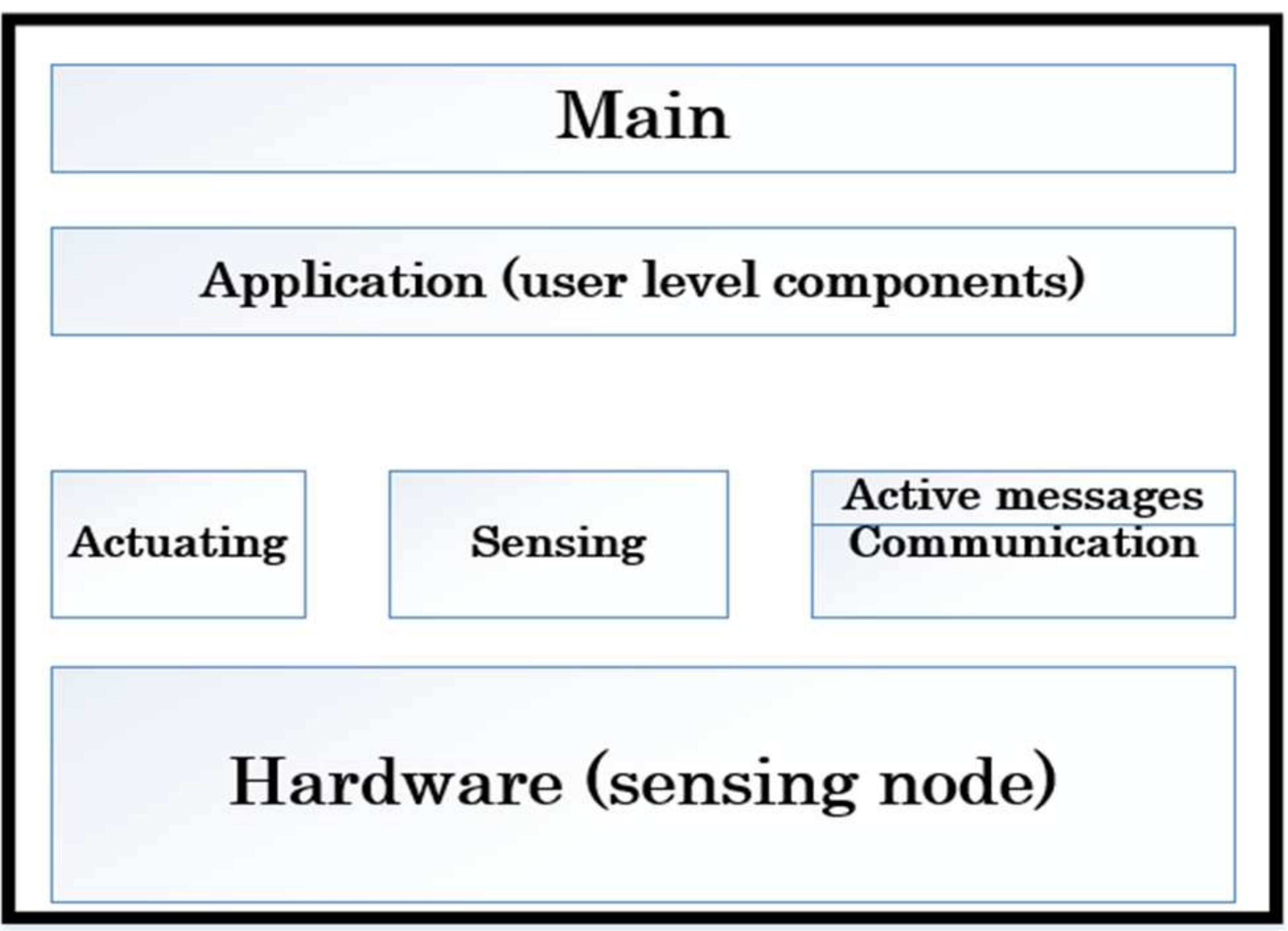}
\caption{Simplified architecture of TinyOS}
\label{fig:alok6-eps-converted-to}
\end{figure}

%
\subsection{MANTIS}
It is a multithreaded operating system used for WSNs. MultimodAl System for NeTworks of In-Situ (MANTIS), is a light weight and energy efficient operating system. The key feature of MANTIS OS is that we can port it across multiple platforms . It is written in C and supports application developed in C. It follows a layered architecture design which is shown in figure 5. The layered architecture of MANTIS OS provides service to the layers above. MANTIS OS supports a multitasking programming model which is pre-emptive \cite{von2003events}. To achieve a better memory management the designer of MANTIS OS logically divided the RAM in two sections \cite{farooq2011operating}, memory for global variables which is allocated at compile time and the rest of the RAM is controlled as a heap. A stack space is allocated when a thread is created from the heap and this stack memory is free when the thread exits. MANTIS keeps the information of threads in a table managed by the kernel. The memory allocation is static for threads so it only supports a limited no of threads. The table contains the information regarding the stack boundary, the current stack pointer, size, priority level of threads, pointer to thread starting function and address of the current point to the next thread address. A MANTIS OS uses a Unix like scheduler, which further supports the functionality of pre-emptive priority based scheduling algorithm. The scheduler in MANTIS uses a Round-Robin scheduling algorithm. The default time slice is set to 10 milliseconds however, it is configurable. Timer interrupts are used by the scheduler for context switching mechanism. The scheduler used in MANTIS is also manages energy efficiency by switching the microcontroller into sleep mode when the threads are idle and not in running state. In this OS, dynamic memory management scheme is followed at the cost of overheads. However, it does not have any memory protection mechanism. The communication protocols supported by the MANTIS OS make it flexible and provide the facility to implement customized routing protocol and transport layer protocols above the MAC layer. Hence, it allows implementing real time transport and routing protocols for WSN which can also be used in multimedia wireless sensor networks \cite{farooq2011operating}.
\begin{figure}[t]
\centering
\includegraphics[width=0.75\linewidth, height=0.3\textheight]{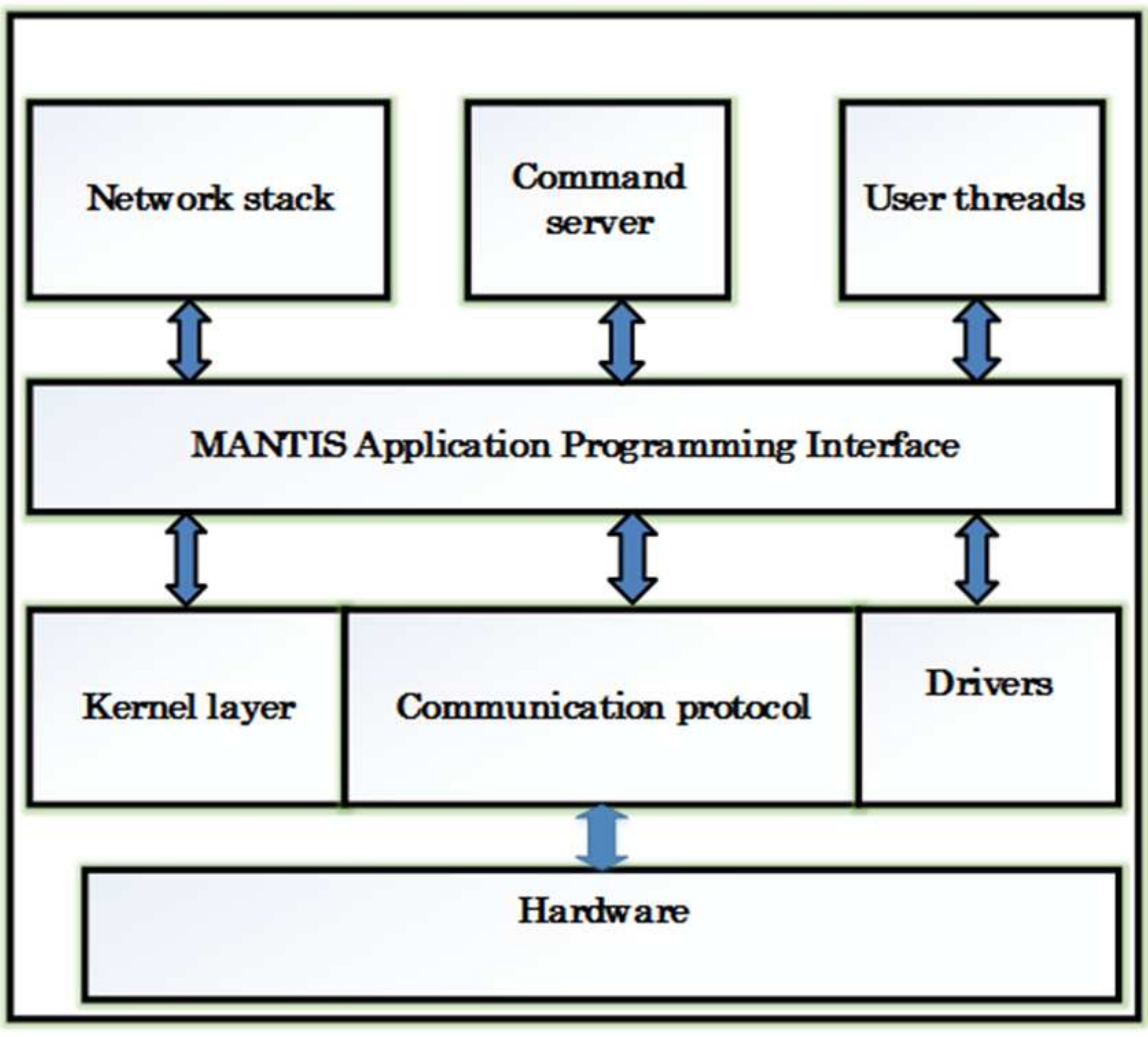}
\caption {MANTIS OS architecture}
\label{fig:alok7-eps-converted-to}
\end{figure}

\subsection{NANO-RK}
Nano-RK \cite{1563113}, is a real time pre-emptive multitasking and fixed OS for WSNs. The main design objective of this OS is to achieve support for multi-hop networking, priority based scheduling, timeliness, small foot print and limited use of resources. It has 2kb of RAM and 18kb of ROM. Nano-RK OS follows a monolithic kernel architecture model. It has as Task Control Block (TCB) and this TCB is initialized during the system image creation and initialization. TCB deals with the information, including priority list, reservation sizes for sensors, CPU, network, resources period and port identifiers of the process. To have a better control, Nano-RK further maintains two linked lists of TCB pointers to order the set of suspended and active tasks. Resource sharing is achieved using reservations of CPU, sensors and network bandwidth. Nano-Rk allows implementing the priority ceiling algorithm. It has a static memory management scheme and there is no dynamic memory management support. The Nano- RK OS has a lightweight protocol stack which provides the communication abstraction similar with the sockets. In this OS, the memory is managed by applications. The concept behind this is that the developers assumed that if an OS reserves memory which is share for only receiving and transmitting data then it is wastage of memory. So, they designed the OS in such a way that the application itself manages its buffer which take care of memory for receiving and transmitting few data bytes. The advantage of such design is that the new incoming data is not entertained unless until the previous data is read by the application or it allows reading the old data.
\subsection{LiteOS}
LiteOS \cite{cao2008liteos}, is developed at the university of Illinois at Urbana-Champaign. It is a Unix-like operating system used for WSNs. The objective of designing this OS is to provide an ease to system level programmers and experience Unix-like environment and support for object oriented programming. It uses Lite C++ as a programming language and a Unix like shell. The LiteOS is used to run MicaZ motes having 128 bytes of flash memory, 4kbytes of RAM and 8 MHz CPU. LiteOS is designed by following modular architecture. It supports communication protocols in terms of file based communication. It has mainly three components: Lite Shell, Lite FS and kernel. Lite Shell deals with the different command support for shell, including debugging, devices, file management and process management. This Lite Shell resides on a PC or at the base station. The only concern with Lite Shell is that it can be used only when a user is present at the base station. Lite FS is responsible for integrating all neighboring sensor nodes and treat them as a file. This further mounts the sensor nodes as a directory and then lists all those nodes which are one hop sensor nodes as a file. This design allows the user at the base station to use this directory as like UNIX directory. A Lite OS kernel resides on the sensor node. A kernel supports multithreading, dynamic loading and uses a round robin scheduling algorithm. A programmer can use callback functions to register event handlers and this further supports synchronization. The LiteOS treats every application as a single thread. This reduces the semantic errors that may occur during write and read operation in the shared memory address. Furthermore, LiteOS provides a mechanism to avoid the race conditions using functions atomic\_start() and atomic\_end() and supports dynamic memory allocation through the use of C-like malloc() and free() functions \cite{farooq2011operating}.
\begin{figure*}[t]
\centering
\includegraphics[width=0.8\linewidth, height=0.4\textheight]{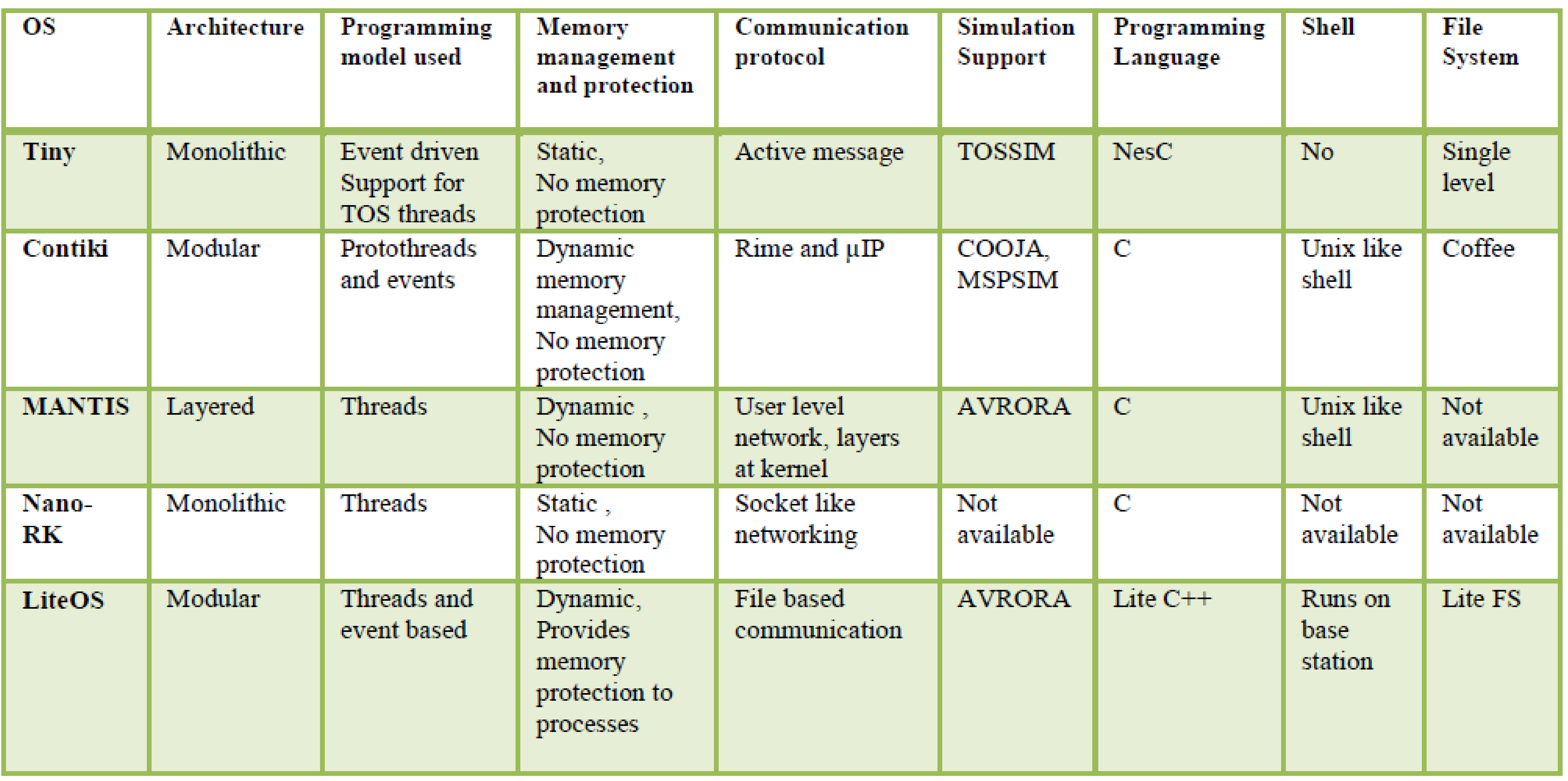}
\caption {comparative study of different OS}

\label{fig:table8-eps-converted-to}
\end{figure*}
\begin{figure}[h]
\centering
\includegraphics[width=0.7\linewidth, height=0.25\textheight]{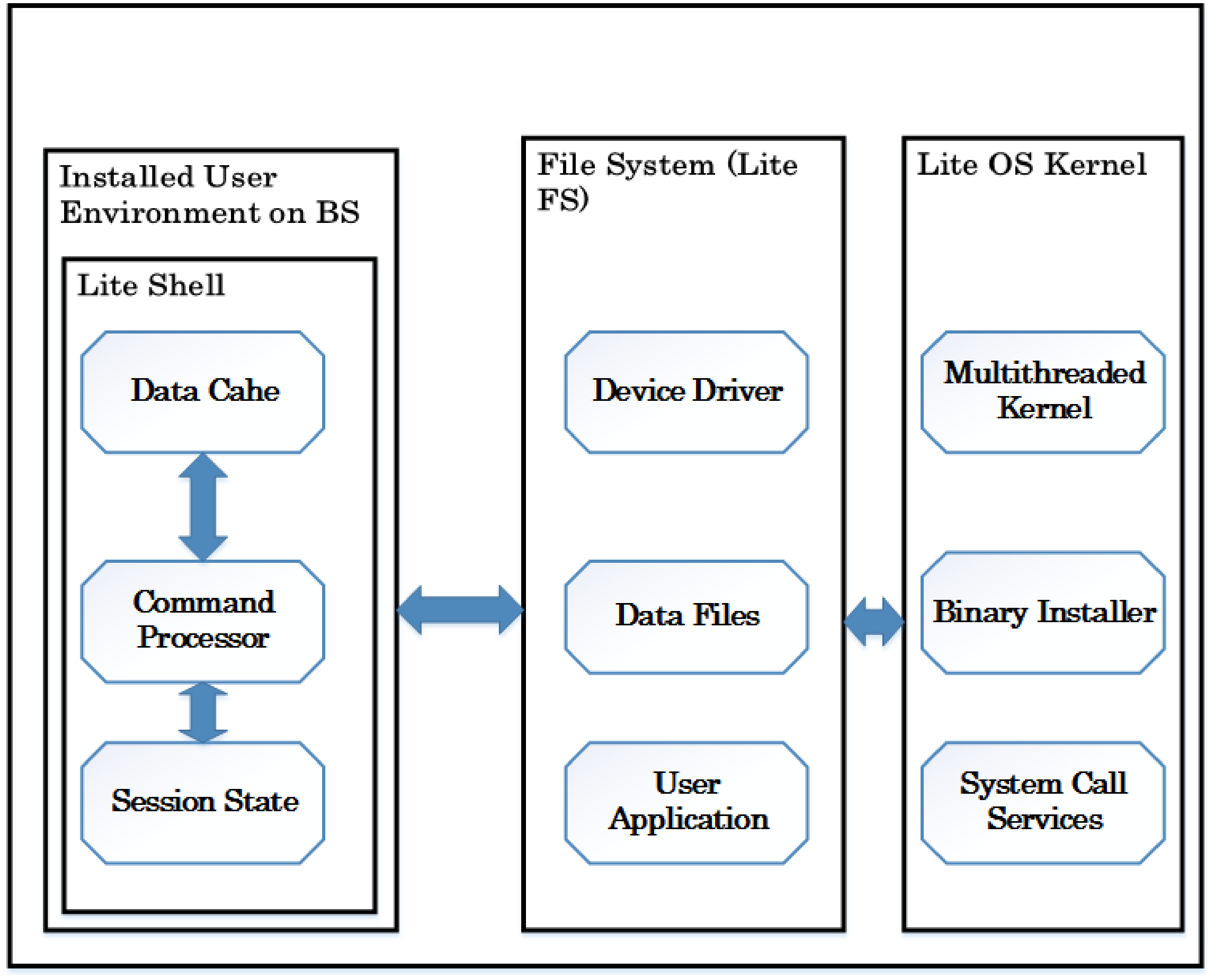}
\caption{Lite OS architecture}
\label{fig:alok10-eps-converted-to}
\end{figure}
\subsection{Contiki}
Contiki OS \cite{1367266}, is a lightweight and flexible operating system which was developed at Swedish Institute of Computer Science (SICS), Sweden. It is developed to target resource constrained devices such as wireless sensor networks. Contiki is implemented in C and is event driven. Although the kernel in Contiki is event driven, but the system supports preemptive multithreading programming module. Preemptive multithreading programming module is implemented as a library in the system and is linked to that program only which need multithreading. In Contiki OS the communication mechanism between processes always goes through the kernel of the OS only. The kernel allows device drivers and applications to communicate directly with the hardware of the system. There are two partitions in Contiki system: the core and loaded program as shown in figure 8. A core in Contiki consists of a kernel, program loader, language run time, communication stacks and support libraries. The core of the system is compiled into a single binary image and generally it is not modified after deployment. The program loader is responsible for loading programs into the systems. The Program loader uses communication stacks for obtaining program binaries. A Contiki OS provides dynamic memory management, but it does not support memory protection mechanisms. Contiki OS implemented communication as a service to provide run time replacement .This gives the advantage of simultaneously loading of multiple communication stacks. Contiki OS has a footprint greater than TinyOS but lesser than MANTIS.
\begin{figure}[h]
\centering
\includegraphics[width=0.7\linewidth, height=0.25\textheight]{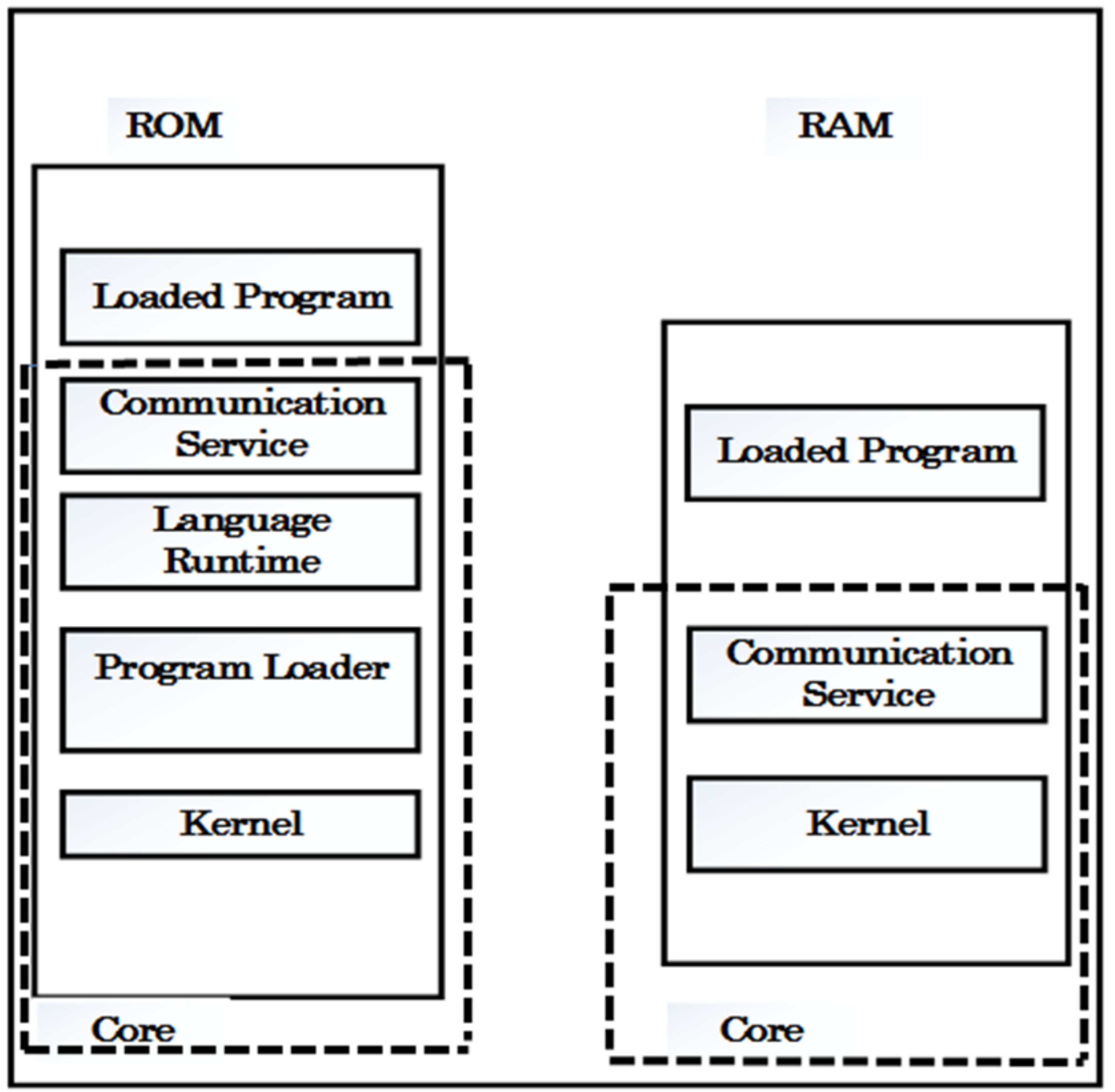}
\caption{Partioning of programs in Contiki OS}
\label{fig:alok10-eps-converted-to}
\end{figure}


\section{Issues with OS for WSN }
With the application needs the design of platforms used in WSN must address parameters like energy efficiency, delay, reliability, cost and scalability \cite{dutta2012operating}. In paper \cite{hui2004dynamic}, the authors have classified the operation of WSN at two levels, which are at the node level and network level. A network level deals with the routing, communication channel, protocols, connectivity, and etc. Whereas, hardware, sensors, energy consumption, radio and central processing unit (CPU) is taken care at the node level. The concern here is efficient resource management and support for hetrogeneity and scalability. To the above said issues, research efforts in both hardware and software are required. Hardware typically includes lower cost while the software part includes network lifetime, middleware, robustness, scalability, reliability and security. To the best knowledge of the authors, there is no any single platform available which can address the variety of application requirements. In the following section, we present open design issues for OS, which are required for reliable performance of WSN based systems. We present issues, including parameter architecture, programming model, scheduling of process, memory protection and support for communication protocols.
\subsubsection{Architecture }
In \cite{phani2007operating}, the authors reported that the services provided by the OS can be influenced by the kernel's architecture which are (a) size of the core kernel and (b) runtime reconfigurability.  In addition to this, an architecture is also responsible for adding services and updating the services to the kernel.  Major architectures which are known and used in OSs are layered architecture, virtual machine architecture, monolithic architecture and the micro- kernel architecture \cite{levis2005tinyos,1563113,cao2008liteos,1367266}. A monolithic architecture does not have any specific structure therefore the services provided by the OS are taken separately or individually. This provides to integrate all services together into a single system image, thus this architecture has very small OS memory footprint. In microkernel architecture the size of the kernel is reduced because the functionality is provided in the kernel. This architecture has better reliability and ease of customizations. The major demerit of this architecture is the poor performance due to frequent user to kernel boundary switching. With the purpose of exporting OS to the user programs, virtual machine architecture is another choice for OS design. The advantage of this system is the portability facilities.  Layered architecture has advantages like easy to manage, simple and reliable in performance but has the major drawback that this architecture is not so flexible. The design considerations for OS should include flexibility, less kernel size and future extension supports. Merits and demerits of different architecture used in OS design are listed in table II.
 \begin{table}%
 \caption{Merits and Demerits of different architecture design\label{tab:table5-eps-converted-to}}
 {\includegraphics[scale=.075]{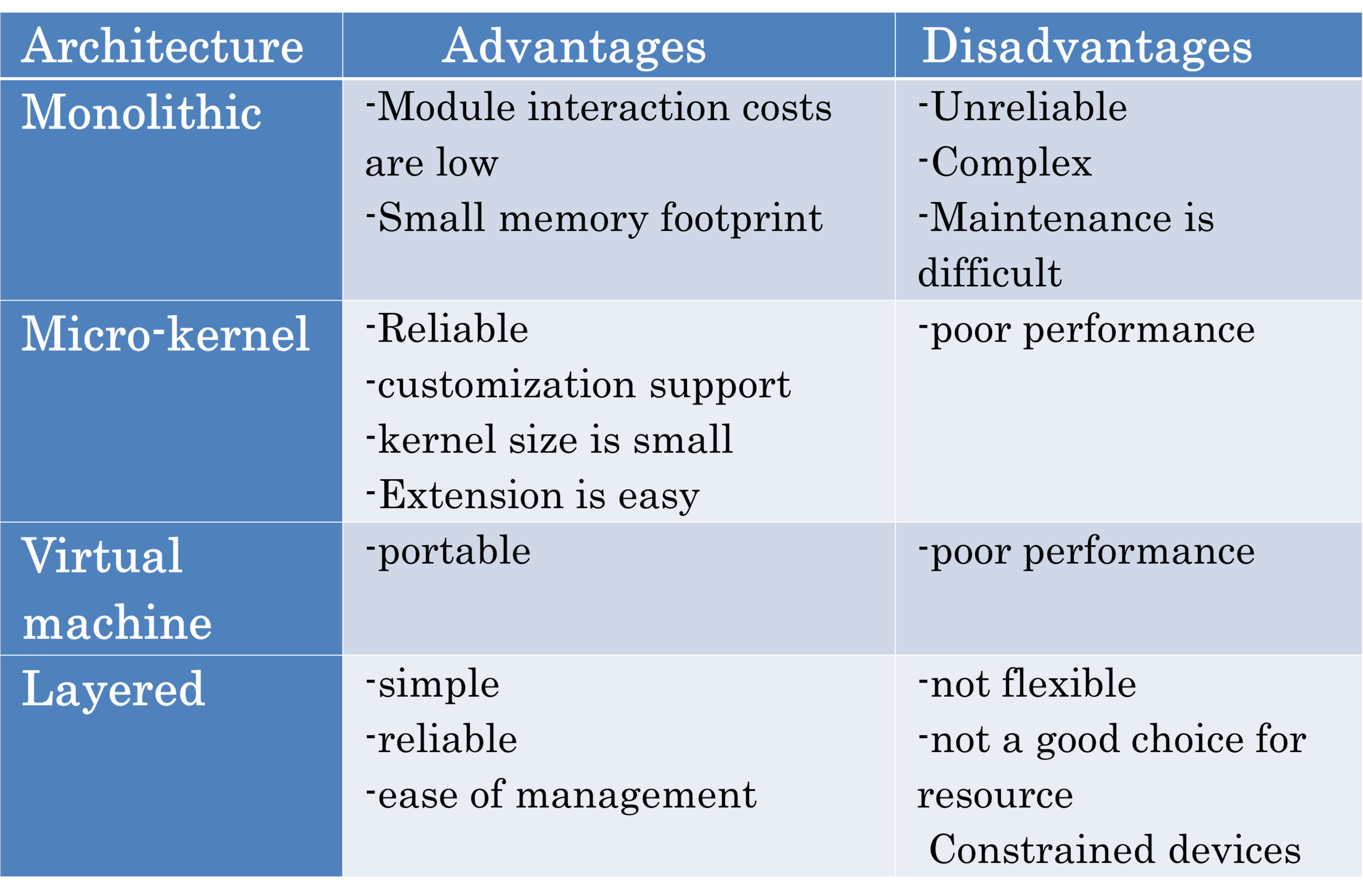}}
 \end{table}

\subsubsection{Programming Model}
A better application programming interface (API) provides a way to have a clear understanding of separation between the application program and the low level node functionalities. An OS should have a clear set of APIs to interact with the developers. In addition to this, the OS API may include sensor data reading APIs, networking APIs, manupulation of memory APIs, management of power APIs and task management APIs \cite{1563113}. This helps in a optimized resource management. The role of a programming model supported by an OS has significant importance for applications. Mainly there are two programming models which are current state of arts for WSN OSs, multithreaded programming model and event driven programming model \cite{farooq2011operating}. A programmer is comfortable with the multithreaded programme and is also used for many application development. But, this is not a good choice for resource constrained devices or networks. Whereas an event driven programming model is useful for computing devices where the role of a resource is critical. But, for traditional applications this has also limited choices. To overcome this researchers have worked and developed a light weight multithreading programming model for WSNs. Reprogramming of an OS is a kind of mandatory feature which can not be avoided. Dynamically updating the software is achieved using this feature only. As, the WSN involves inaccessible deployment, reprogramming of the platform is critical and to do this dissemination protocol is used to distribute the code \cite{hui2004dynamic,stathopoulos2003remote}. For a successful reprogramming the code should be relocatable and can be run in any memory location.
\subsubsection{Process Scheduling}
To avoid race conditions, an efficient synchronization mechanism is needed in the OS. A flexible computational support aids in design of the flexible architecture of the OS. In the case of a priority application, scheduling of computational units is very much crucial. The sequence or order in which a task should be executed is another important concern while designing OS. A CPU is responsible for scheduling the tasks or processes. As a WSN is widely used in various applications including real time applications and non-real, the importance of an efficient scheduling is felt critical. Researchers are working on scheduling algorithms so that the OS can accommodate the requirements by application \cite{levis2012experiences}.
\subsubsection{Memory Protection and Memory Management}
Memory protection refers to protect one process address space from another, so that there should not be any conflict in process execution. A WSN has very limited memory for operation. A better memory management can significantly improve the performance of the network. Therefore, the design for OS should have a good memory management scheme.  Earlier OS designers for WSN assumed that only one application executes on a sensor node therefore memory protection mechanism is not required\cite{farooq2011operating}. But in recent years' application requirement, it is desired to have the memory protection scheme \cite{dutta2012operating}.
\subsubsection{Support for Communication Protocol}
Communication in the OS design context refers to the process in which a device can not only share data within the system itself but also with other nodes in the network. A WSN is distributed in nature and the nodes communicate with each other in the network. With the future prospects application demand, to support heterogeneity and interoperability features a communication protocol should be developed with the considerations of one system to all. Research efforts are needed to support the inter component communication which further helps in dynamic linking of the components \cite{stathopoulos2003remote}.
\section{Future Remarks}
The OS research community has done extensive work on the reliability of the OS for WSNs. However, there are still open research issues available with the consideration of new application areas and associated challenges to the WSNs. The architecture of an OS influences the size of the kernel and also has an influence on the mechanism to provide the services to the application programs. Resource management is the elementary problem for the operating system. The role of an OS is very crucial in terms of the resource management. The limited resources of sensor nodes in terms of power, computation, reliability are some challenges which need to be considered while designing the operating system. The interface provided by the OS should be easy to use for the developer. This results in addition set of challenges to the way the OS can be developed by the designer. Recently, the application area of the tiny networked devices is increasing rapidly  such as multimedia data processing, video surveillance in a hostile environment, industry automation, underground mining communication and sensing and  underwater communication. The real time application support is available in few OS. In future research, smart algorithms which may support these application areas that can accomodate hard and soft real time requirements of applications will significantly help in the reliable communication. As the future application of WSN is increasing day by day; it is worth to mention here, that in future large memory space would be a critical issue with the OS. Therefore, memory management and protection and secondary storage support is another research area to be explored. This may further require efforts in the area of virtual memory management.  A WSN may have an application area where accessibility of the deployed sensor node is not frequent and in some cases not possible like volcano monitoring and nuclear reactor plant. In such cases, multi application support is needed. For example, a sensor node may have different objectives of deployment like capturing humidity, temperature and sensing the vibration. Therefore, an OS should accommodate multiple applications at the same time. Due to the limited resources, unattended deployment, high dynamic behavior of topology in WSNs, the design approach of OS for tiny networked sensors deviates from the traditional OS \cite{phani2009operating}. With the vision of the future application demand of tiny network, an OS should be designed in such a way that portability of the OS may possible at minimal changes in hardware. This is because, every application designer is working on the customized hardware platforms as per their application needs. It is also important that an OS should be capable of network dynamics because of the mobility issues in WSNs and failure of the nodes or communication links in the network. Therefore, the OS should adapt the dynamic changes in topology.

 \section{Conclusions}
 In this paper, we have surveyed the popular known OS widely used for WSNs. We also discussed the open research issues of communication protocols followed by comparative study on the different OSs. Different design issues of an OS design  are also discussed. We believe  that, this survey paper may help the designer research community to design more reliable and robust OS for tiny networks.

  \section*{Acknowledgment}

The authors would like to thank Director, NIT Rourkela for his motivations to carry such research.

\ifCLASSOPTIONcaptionsoff
  \newpage
\fi


\bibliographystyle{IEEEtran}
\bibliography{alokref}

\end{document}